# NMR molecular photography

**Anatoly K. Khitrin, Vladimir L. Ermakov[a], and B. M. Fung[b]**

*Department of Chemistry and Biochemistry, University of Oklahoma, Norman, Oklahoma 73019-3051*

*Abstract*

A procedure is described for storing a 2D pattern consisting of $32\times32 = 1024$ bits in a spin state of a molecular system and then retrieving the stored information as a stack of NMR spectra. The system used is a nematic liquid crystal, the protons of which act as spin clusters with strong intramolecular interactions. The technique used is a programmable multi-frequency irradiation with low amplitude. When it is applied to the liquid crystal, a large number of coherent long-lived $^1$H response signals can be excited, resulting in a spectrum showing many sharp peaks with controllable frequencies and amplitudes. The spectral resolution is enhanced by using a second weak pulse with a 90° phase shift, so that the 1024 bits of information can be retrieved as a set of well-resolved pseudo-2D spectra reproducing the input pattern.

[a] On leave from Kazan Physical-Technical Institute, Kazan 420029, Russia.
E-mail: ermakov@ou.edu (current); ermakov@smtp.ru (permanent).
[b] To whom correspondence should be addressed. E-mail: bmfung@ou.edu.

Electronic information storage and processing are an indispensable part of modern technology. At the present time, devices for the storage and processing of information are largely based on semiconductors. Further technological advances depend heavily on the reduction of the sizes of these devices, ultimately to the molecular level. For examples, molecular wires, rectifiers, and semiconductors[1-6] and other molecular electronic devices such as switches, sensors, and storages[7,8] have been described. Many molecular information storage devices are based on compounds with complicated structures requiring elaborate syntheses,[9,10] but the number of bits per molecule is very small.

In this report, we describe a novel and exciting method for storing and retrieving a 2D pattern in a molecule. It is based upon a completely different system, namely clusters of nuclear spins. The input information is programmed in the form of 1K (32×32) "pixels", and retrieved as a set of pseudo-2D NMR spectra reproducing the input pattern. The material used is a commercially available liquid crystal, and no special synthesis is required.

Liquid crystal devices (LCD's) are widely used for macroscopic electronic information storage and processing. We have demonstrated that it is also possible to store information in individual liquid crystal molecules.[11] Here we show how to use liquid crystal molecules to store and retrieve 2D patterns, a process similar to photography. The use of NMR for information storage was suggested many years ago for isotropic liquids.[12,13] When the sample is placed in a magnetic field with field gradient, the use of a weak radiofrequency (rf) field can excite a narrow band of the spectrum of the inhomogeneously broadened signal. The shape of the weak excitation pulse can be imprinted on the response signal in the form of spin echoes,[12] and this phenomenon can be used for information storage.[12,13] However, the information is stored in different parts of a macroscopic sample rather than at the molecular level, and the process is sequential, which limits the content of input information. In contrast, the input is parallel in our method, and the information is stored in individual molecules.

The potential power of using spin clusters for molecular information storage and processing is based upon the high dimensionality of the associated Hilbert spaces: a system of $n$ coupled spins (with I = ½) can have up to $C_{2n}^{n+1} \sim 2^{2n}$ peaks in its NMR spectrum. To gain access to these high dimensions and the informational capacity they offer, the spins must interact with each other to create correlated states. However, it is very difficult to use conventional NMR techniques to manipulate individual peaks without affecting the others for the purpose of controllable information storage.

As a desirable step towards high-content molecular information processing, one needs to learn how to encode and read the large amount of information that can be stored in the states of a system of coupled spins. For this purpose, we have explored the use of a new technique applied to systems whose NMR spectra do not show resolved peaks under normal conditions.[11]

A liquid crystal molecule is a cluster of spins coupled by dipole-dipole interactions, and its $^1$H NMR spectrum is very broad. For example, in the nematic liquid crystal 4'-*n*-pentyl-4-biphenylcarbonitrile (5CB), the 19 proton spins of each molecule are coupled with dipole-dipole interactions of various magnitudes, and the spectrum obtained by applying a non-selective strong pulse shows an ill-resolved broad signal because of extensive overlapping of numerous peaks (Fig. 1A). Its width of about 25 kHz is determined by the residual dipole-dipole interactions between the 19 proton spins of the molecule due to orientational order, while intermolecular dipolar couplings are averaged to zero due to rapid translational motion. Many sharp peaks can be observed with an acquisition delay,[14] but it is virtually impossible to manipulate them for the purpose of information storage.

Using a different approach, we have found that a sharp coherent signal (linewidth ≈ 12 Hz) can be observed when a weak rf pulse is applied to the spin cluster.[11] The phase of the peak is opposite to that of the conventional NMR spectrum, and part of the broad component is also excited. The origin of the sharp peak is entirely different from the sharp peaks obtained using a strong rf field with acquisition delay, and a computer simulation of a small spin cluster does reproduce the basic features of its response to a long and weak pulse.[11] Furthermore, many narrow peaks at the individual excitation frequencies can be observed simultaneously by applying a multi-frequency irradiation (Fig. 1B). The number, positions, and magnitudes of these sharp peaks can be completely controlled by programming the weak multi-frequency rf excitation.

Because versatile multi-frequency excitations can be achieved by modulating the amplitudes and phases of rf pulses to create multi-harmonic pulses,[11,15-17] it is possible to excite a large number of sharp peaks for molecular information storage. By setting some amplitudes of the harmonics in the excitation pulse to zero and the rest to a fixed value, one can encode a bit string, and then retrieve this information from the NMR spectrum.[11] The actual density of information stored and retrieved is restricted by the width of the coherent peaks and, even more importantly, the overlapping of the broad components. For 5CB, the resolution limit is about 250 peaks. However, by using an additional pulse, recording with a much higher density can be resolved. The principle and the procedure of the technique are explained in the following.

In the two-pulse experiment, the second weak pulse is phase-shifted by ±90° with respect to the first one, similar to a spin-locking experiment. When phase cycling is used to eliminate the direct signal produced by the second pulse, the amplitude of the resulting signal shows a very unusual dependence on the length of the second pulse (Fig. 2A): as the length of the second pulse increases, the signal changes sign and can even be amplified. This result is rather unexpected, highly intriguing, and somewhat counter-intuitive. However, a theoretical analysis or even a complete numerical simulation is impossible because the spin dynamics of a coupled 19-spin system in its huge Hilbert space is extremely complicated. Therefore, we performed a computer simulation of a much smaller (8-spin) system to ascertain that the observed phenomenon is not an experimental artifact. The results are shown in Fig. 2B. Because of the reduction in the number of coupled spins, the simulated results do not agree completely with the experimental data. Nevertheless, the essential features, namely the reversal of the sign and the amplification of the signal as the length of the second pulse is increased, are reproduced in the simulation.

The lifetime of the signal produced by applying two consecutive weak pulses, the first one for excitation and the second one for "locking," lasts considerably longer than the dephasing rate of 12 Hz for the coherent signal produced by a single excitation pulse. Therefore, one can create a situation where an appreciable signal is detected only when both pulses have non-zero amplitudes for the corresponding harmonic, effectively enhancing the resolution by a factor of about 4, from 250 peaks to over 1000. As an example, an image consisting of 32×32 = 1024 bits is recorded in a spin state of 5CB molecule and displayed in Fig. 3. It can be seen that the 1 K image is reproduced with good fidelity in the NMR spectra. We call this technique "NMR molecular photography", in which the dark and white elements of the image (ones and zeros, respectively) are transformed into a set of NMR spectra by using the following procedure.

Basically, the whole information consisting of all "pixels" is completely imprinted in the spin cluster by the first pulse but retrieved one row at a time by using the second pulse in a quasi-two-dimensional experiment. To do this, the input image is represented as a one-dimensional 1024-bit array of zeros and ones starting from the left upper corner, column after column. Then, the first pulse is programmed correspondingly as a sum of 1024 circularly polarized harmonics: the amplitudes of the "zero" bits are set to zero, while the amplitudes of the rest of the harmonics are all equal and correspond to a precession frequency ($\gamma B_1/2\pi$) of 1.2 Hz. For this multi-frequency pulse, successive frequencies, in steps of 20 Hz each, are assigned to every bit, so that the total spectral range is 20×1023 = 20460 Hz. The second pulse is designed to retrieve part of the stored information, namely only one row in the image each time. It has 32 harmonics with equal amplitudes (precession frequency $\gamma B_1/2\pi$ = 9.0 Hz), and the frequency interval between the harmonics is set to 20×32 = 640 Hz. When the reference frequency of the second pulse is increased by 20 Hz, the next row of the image is reproduced. By shifting the frequency 31 times, the restored image is obtained as a stack of NMR spectra (Fig. 3), in which the higher peaks correspond to the dark cells of the input image. Experimentally, in order to practically implement each multi-frequency pulse, the amplitude and the phase of the rf field is modulated in a complex way by sampling the ideal waveforms at a constant rate. The first pulse with 1024 harmonics consists of 50 K elementary steps, and has a duration of 1.0 s; the second pulse with 32 harmonics consists of 10 K elementary steps, and has a duration of 0.05 s.

Since the interactions between different molecules are averaged by fast molecular motions, the total spin Hamiltonian of the system is a sum of commuting Hamiltonians $H_i$ for individual molecules: $H = \Sigma H_i$. Therefore, the initial density matrix is a product of identical spin density matrices for individual molecules:

$$\rho_o = Z^{-1} exp(-\Sigma H_i/kT) = Z^{-1} \Pi \, exp(-H_i/kT) = \Pi \, \rho_{oi}, \tag{1}$$

where $Z = Tr\{exp(-\Sigma H_i/kT)\}$. It can be seen that evolution caused by spatially homogeneous fields does not change this property, and the spin density matrix at all times is a product of identical density matrices for individual molecules. Therefore, the information is recorded on a single spin cluster, and the other molecules of the sample contain identical copies of the same information. The use of a macroscopic sample was necessary for overcoming the inherently low sensitivity in NMR experiments and suppressing quantum fluctuations, but the sample size can probably be greatly reduced for electron spin systems.

Two other remarks need to be made. First, the storage of information using spin clusters is not permanent. From Fig. 2 one can see that the "locked" signal decays with the characteristic time of about 0.1 s, which is sufficient for processing the stored information. Second, using the terminology of logic gates, the application of the second pulse is an implementation of parallel bitwise AND operation. An example of parallel NOT operation is given elsewhere.[18]

Excitation of long-lived coherent signals in dipolar-coupled systems with unresolved spectra is a very interesting phenomenon. We hope that the results presented here will stimulate further investigations, including theoretical studies of the spin dynamics of complex systems.

ACKNOWLEDGMENT

This work was supported by the National Science Foundation under grant numbers DMR-9809555 and DMR-0090218.

**Figure Captions**

FIG. 1. $^1$H NMR spectra of 5CB at 400 MHz and 20 °C. (A) Spectrum obtained by applying a 90° pulse with 5 μs duration, and (B) spectrum obtained by applying a 0.2 s pulse which is the sum of 10 circularly polarized harmonics, each having an rf field amplitude of 12 Hz. The spectrum in (B) is shown with a 10-fold magnification.

FIG. 2. (A) Experimental dependence of the $^1$H signal amplitude of 5CB on the duration of the second pulse in a two-pulse experiment. A long-lived coherent $^1$H signal with a linewidth of *ca.* 12 Hz was created by applying a 1 s single-frequency pulse with $\gamma B_1/2\pi = 2.1$ Hz; then, a second pulse with $\gamma B_2/2\pi = 11.0$ Hz and a 90° phase shift was applied. The direct signal produced by the second pulse was removed by phase cycling. The number of transients was 64 for each of the spectra. (B) Computer-simulated spectra obtained by the two-pulse experiment for a model 8-spin system. The system used has random dipolar couplings ranging from −792 to 792 Hz; the duration of the first pulse is 1.0 s with an rf amplitude of 1.0 Hz; the duration of the second pulse is varied from 0 to 0.20 s in steps of 0.01 s, and its amplitude is 4.0 Hz. A line bradening factor of 12 Hz is used. Actually the frequency of each peak is the same in the 21 simulations, but each spectrum is shifted successively by an increment of 400 Hz and the results are superimposed on each other, so that a visual comparison with the experimental data shown in (A) is more straightforward.

FIG. 3. "Molecular photography" showing a 32×32 = 1024-bit image (top) imprinted on the proton nuclear spin cluster in the 5CB molecule and then retrieved as a stack of $^1$H NMR spectra (bottom). A consecutive two-pulse sequence was applied to obtain each spectrum. The first pulse had a duration of 1.0 s; each of its 1024 harmonics had a precession frequency ($\gamma B_1/2\pi$) of 1.2 Hz. The second pulse in each row had a duration of 0.05 s; it was phase-shifted by ±90° with phase cycling; each of its 32 harmonics had a precession frequency ($\gamma B_1/2\pi$) of 9.0 Hz. The number of transients was 512.

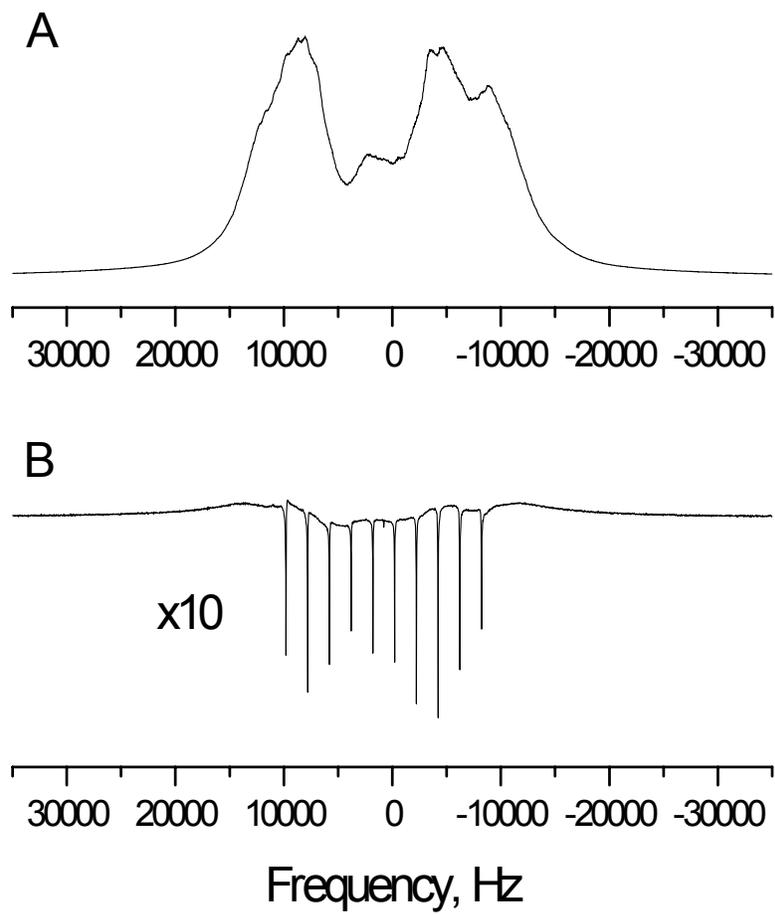

Fig. 1

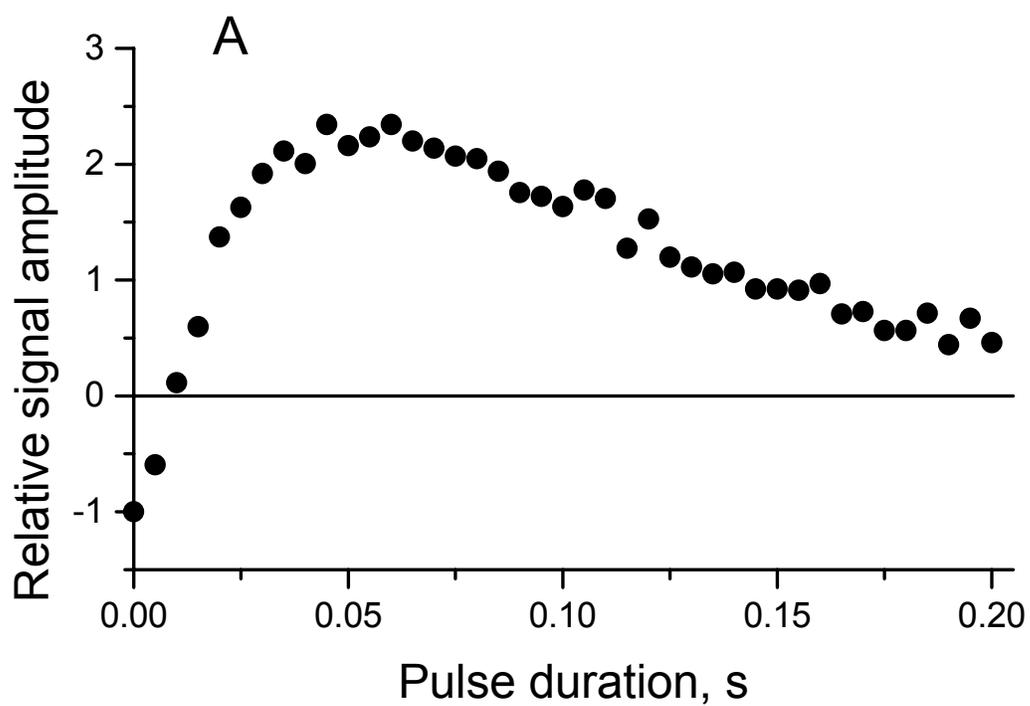

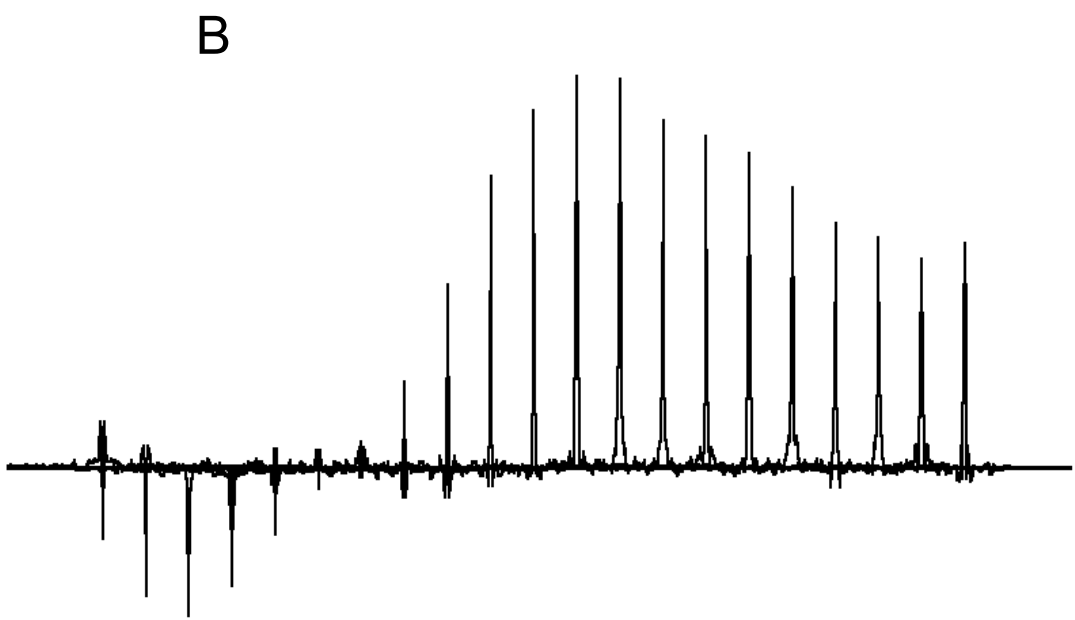

Fig. 2

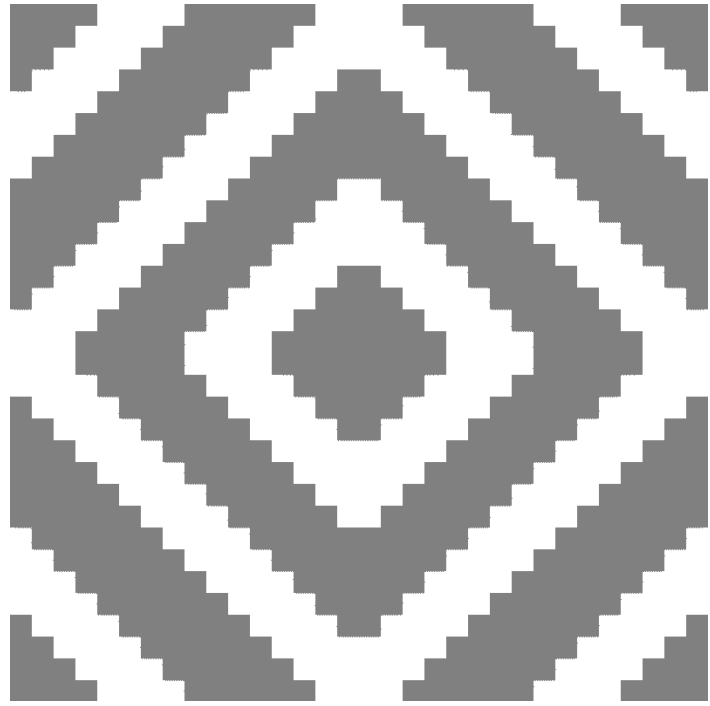
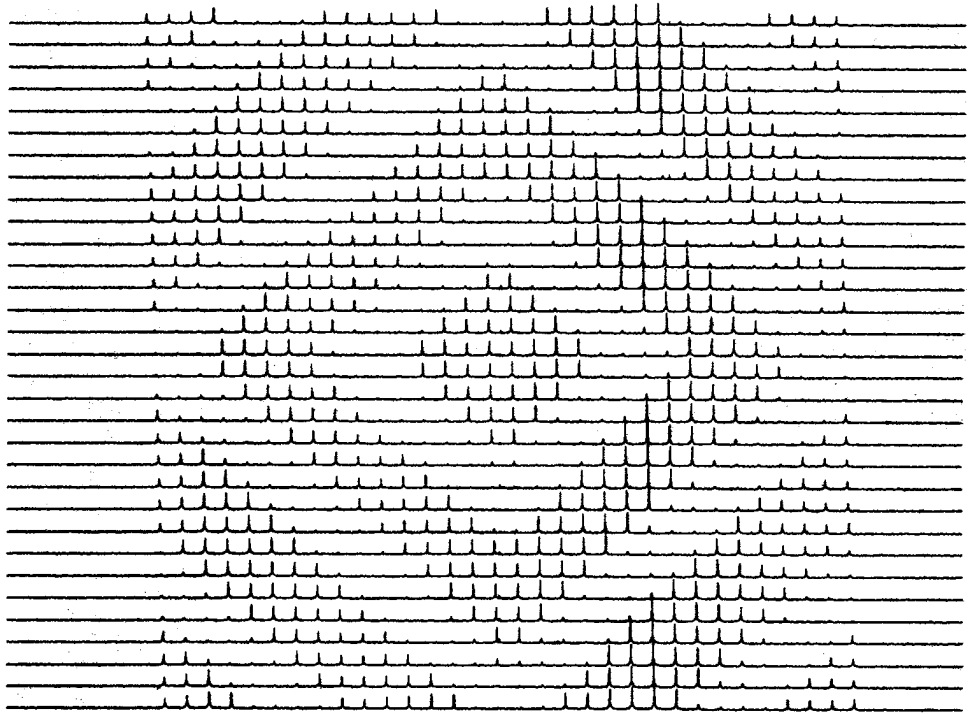

Fig. 3